\documentclass[pra]{revtex4}
\usepackage{epsfig}
\usepackage{bm}

\begin{document}

\title{Probabilistic purification of noisy coherent states}
            
\author{Petr Marek}

\affiliation{School of Mathematics and Physics, The Queen's University,
  Belfast BT7 1NN, United Kingdom}

\author{Radim Filip}
\affiliation{Department of Optics, Palack\'y University, 17. listopadu 50, 77200 Olomouc, 
Czech Republic}

\affiliation{Institut f\"{u}r Optik, Information und Photonik,  Max-Planck Forschungsgruppe, Universit\"{a}t Erlangen-N\"{u}rnberg, G\"{u}nther-Scharowsky-Str. 1, 91058, Erlangen, Germany}

\begin{abstract}
A basic feasible probabilistic purification of unknown noisy coherent states, outgoing from
different state preparations with unknown mean number of thermal
photons, is proposed. The scheme is based only on a linear-optical
network with an avalanche photo-diode or heterodyne (homodyne)
detection used to post-select a successful processing. The suggested
probabilistic method can produce an output state with a lower noise
than both quantum deterministic and classical probabilistic
distillation method. The purification applied in the state
preparation can increase classical capacity of communication and
security of quantum key distribution.
\end{abstract}
\maketitle

\newcommand{\dd}{\mathrm{d}}
\newcommand{\ee}{\mathrm{e}}
\newcommand{\ii}{\mathrm{i}}

\section{Introduction:}          

Quantum information processing with continuous variables (CVs)
alternatively renders feasible protocols based previously on
discrete variables \cite{CV}. In CV quantum information processing
with light, continuously modulated coherent states of laser beam
are used as carriers of information and homodyne (heterodyne)
detection is employed in measuring them. Ideally, such coherent
communication schemes offer large information capacity and
high transmission rate. However, excess noise present in the
coherent states carrying information can substantially reduce the
classical capacity \cite{capacity,Hall} and break the security of
quantum key distribution \cite{CVcrypt}. Finding a way to reduce
the noise added to coherent states is then clearly a task
of interest.

Any communication task consists of three basic steps: quantum state
preparation, transmission and detection. Different methods, aimed at
reducing the noise, can be applied to each step. When trying to find
possible methods to reduce the noise, we are practically limited by
our ability to control both a noise and a signal, as well as by
experimental limitations.

In coherent-state quantum key distribution, it is necessary to
reduce noise to ensure unconditional security \cite{CVcrypt}. When
attempting to securely transmit coherent states through a noisy
channel, non-Gaussian CV quantum repeaters (based on entanglement
distillation and entanglement swapping) distributed along channel
can, in principle, be used to produce highly entangled and pure
state shared in between two distant parties, as for qubits
\cite{repeat}. Then, an unknown coherent state can be securely
teleported with a high fidelity, using only a noiseless classical
communication link \cite{telep}. However, CV entanglement
distillation of the Gaussian states is not allowed using Gaussian
local operations and classical communication \cite{nopurif} alone.
Therefore, a hard venture beyond the border of the Gaussian methods
is required; for example, by means of high-order nonlinear
interaction \cite{CVdist} or single-photon subtraction \cite{subth}.
Recently, a promising method proposed the use of single-photon
subtraction to produce many copies of an entangled non-Gaussian
state that are subsequently distilled into a single state with
larger entanglement by a Gaussification process \cite{Gaussif}.

On the other hand, when quantum states are used for classical
communication, the most prevalent method of improvement lies in
quantum error detection and correction. Naturally, in this case, it
is reasonable to assume the lack of an equivalent or better
classical channel, because that would undermine the need for using
quantum states to communicate. Thus the above listed quantum
repeaters, realizing a perfect quantum channel, cannot be used. The
simplest classical error detection schemes employ a redundancy.
Information carried by multiple copies (repetition code) is
transmitted through channels and, afterwards, if obtained values
disagree then error is detected \cite{kahn}. Although perfect
copying of quantum states is not possible, due to the no-cloning
theorem, a single unknown coherent state may be still spread to many
modes, propagating through many channels, and arbitrary erroneous
displacement, occurring randomly only in a single unspecified mode,
can be corrected \cite{CVerror}. However, these methods work only if
in most cases, most of the channels are left undisturbed.

Whereas quantum error detection (correction) is designed to correct
the errors in the transmission of a single quantum state, a
different approach, called {\em quantum purification} of states
carrying information, is more suitable for a noise reduction in the
state preparation \cite{symm,purif}. As opposed to the quantum error
correction approach, quantum purification is not restricted by the
no-cloning theorem, because the quantum-state preparation starts
from a classical signal. Therefore, if we consider coherent state
communication, instead of a single unknown coherent state that
cannot be copied, our input is a classical complex amplitude that
can be used to prepare many copies of the coherent state. A standard
method of preparation of any coherent state is by the amplitude and
phase electro-optical modulation of a laser beam. However, the laser
beam itself exhibits low-frequency excess noise \cite{laser} and
also the electro-optical modulators \cite{modul} are devices
exhibiting excess phase-insensitive or phase-sensitive noise,
especially for high speed and large broadness of modulation
\cite{Lod}. In a combination with a lossy channel, such the noisy
encoding decreases secure key rate and may break security of the
communication \cite{security1, security2}.

Let us assume several imperfect copies of a state, produced in the course of noisy state preparation. 
Now, continuous-variable version of the
symmetrization \cite{symm} (which, for qubits, allowed for average
error reduction by a factor corresponding to number of copies) can
be used to concentrate information from number of noisy copies into
a single state. The sender is attempting to prepare a coherent state
$|\alpha_0\rangle$. The noisy modulators produce $M$ (generally
different) noisy Gaussian states
$\rho_{1}\otimes\ldots\otimes\rho_{M}$ with additive white
phase-insensitive Gaussian noise described by mean number $N_i$
($i=1,..,M$) of thermal photons. The task, the sender is faced
with, is to get (at least probabilistically) a single Gaussian state
$\rho'$ having maximal fidelity $F'=\langle
\alpha_0|\rho'|\alpha_0\rangle$ with the target state, using prior knowledge about
noise present in the preparation. This unity gain fidelity
$F'=1/(1+N')$ is function of mean number $N'$ of thermal
photons after the purification. 

Inspiration for efficient purification methods arises from the
classical data processing. There, information carried by many copies
of an unknown noisy signal can be faithfully concentrated and the
noise can be reduced by (deterministic) data averaging or
(probabilistic) data selection. The deterministic data averaging
always produces an output copy with identical signal, but the noise
is averaged over all incoming copies. On the other hand, in the data
selection method, the output is accepted only if all the copies are
similar, otherwise it is discarded. In a direct quantum extension of
the averaging method, an optimal measurement of non-orthogonal
coherent states introduces a noise penalty. Fortunately, using
quantum interference, the averaging can be performed directly
without any measurement involved. Based on this idea, a quantum
Gaussian purification of noisy coherent states from two identical
channels with a superposed Gaussian additive noise has been recently
experimentally investigated using only linear optics \cite{Ulrik}. A
similar method was also recently tested for single photons
\cite{Ricci}.

The quantum purification method strongly depends on prior knowledge
about multiple noisy state preparations. Even if all of them
exhibit common Gaussian additive phase-insensitive noise, they can
substantially differ in mean number $N_i$ of thermal photons added
to the particular preparation $i$. If the sender has perfect
knowledge about the amount of noise in all the copies, the
deterministic purification \cite{Ulrik} can be adapted to the
situation. However, when dealing with an unstable noise level in the
preparation, this is always based on their ability to do it many
orders faster than the speed with which noise parameters are
changed. Even if the actual amount of noise ($N_i$) in the
particular state ($i$) is unknown, the sender can still learn about
the total statistical distribution of $N_i$ in the state
preparations by an long-time probing. But if the change of $N_i$ is
fully chaotic, there is no stable statistics of $N_i$. In this case,
the sender cannot adapt the purification method to some particular
stable statistics of $N_i$ and must attempt to devise a more
universal scheme improving the state preparation for arbitrary
statistics of $N_i$. This is the task we are interested in, the
purification of coherent states carrying information, without any
knowledge about the amount of disturbing Gaussian noise. An
evaluation of the purification is also limited by this lack of
knowledge, an average noise over any particular statistics of $N_i$
cannot be taken as a figure of merit. We can only compare how the
output mean number $N'$ of thermal photons depends on $N_i$. The
optimal method will lead to the lowest $N'$ for all values of $N_i$
imaginable.

\begin{figure} [htbp]
\centerline{\psfig{file=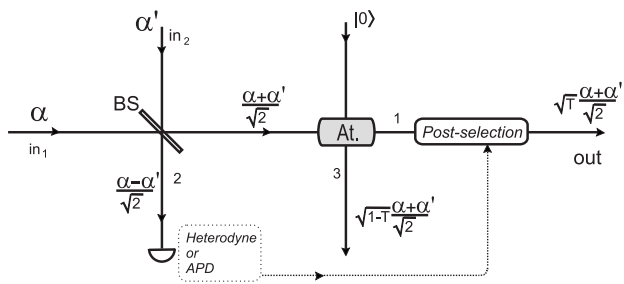, width=8.2cm}} 
\caption{\label{setup} A scheme for probabilistic
purification of coherent states (two-copy purification): BS -- beam
splitter, At. -- attenuator with a parameter $T$.}
\end{figure}

In this paper we present a feasible probabilistic purification
scheme reducing Gaussian additive noise, with unknown mean numbers
of thermal photons, in the state preparation of coherent states. We
show that this method outperforms the deterministic Gaussian
protocol based purely on the mutual interference \cite{Ulrik}. In
addition, the method also beats probabilistic classical methods
based on optimal measurement and preparation. As a figure of merit
in our case of additive Gaussian noise we can simply consider the
mean number of thermal photons after the purification, which is
simply related to fidelity with ideal prepared state. This is also
relevant parameter needed to obtain the capacity of the Gaussian
channel used by Alice and Bob for coherent state communication
\cite{capacity,Hall} and security quantum key distribution
\cite{CVcrypt}. For the classical capacity, a reduction of the
excess noise in the state preparation increases capacity and
similarly, in the key distribution protocol, secure key rate can be
improved. Even the reduction of noise via the purification can
reveal security of the key distribution through lossy channel
\cite{security1}.

The proposed probabilistic purification protocol,
representing the quantum analogue of the classical data selection
method, is feasible, based on using an avalanche
photo-diode or heterodyne detection in the linear-optical setup
and appropriate post-selection of the output state. Approaches
employing avalanche photo-diodes in the CV experiments have been
recently used to produce non-Gaussian statistics from individual
pulses of squeezed light by single-photon subtraction
\cite{subexp}. Single-photon subtraction is a basic element of
many theoretical proposals based on the squeezed states
\cite{subth}. Our scheme is based on a post-selection according to
the detection of the vacuum state by the avalanche photo-diode
(APD), previously used in the Gaussification procedure
\cite{Gaussif} for entanglement distillation.  As an alternative,
we show that heterodyne (homodyne) detection, having substantially
higher efficiency than the APD, can be used in the proposed
purification to achieve a higher fidelity but at the cost of a lower
success rate.

\section{Purification method:}          

In following we will consider the state preparation as a noisy
displacement operation
\begin{equation}\label{Gen00}
\rho = \int \Phi(\beta)\mathrm{D}(\alpha_0+\beta)|0\rangle\langle 0|
\mathrm{D}^{\dag}(\alpha_0+\beta)\dd^2 \beta,
\end{equation}
where $\Phi(\beta)$ is a Gaussian complex probability distribution
of the displacement parameter with zero mean value.
$\mathrm{D}(\gamma)$ stands for the displacement operator
$\mathrm{D}(\gamma)=\exp(\gamma a^{\dag}-\gamma^{*}a)$,
$|\alpha_0+\beta\rangle=\mathrm{D}(\alpha_0+\beta)|0\rangle$,
$|0\rangle$ is the vacuum state and $|\alpha_0+\beta\rangle$ is a
coherent state. After the channel, an input state of our
purification device can be written in the form
\begin{equation}\label{Gen01}
    \rho = \int \Phi(\beta)|\beta+\alpha_0\rangle\langle\beta+\alpha_0|\dd^2\beta.
\end{equation}
Now, having only several (at least two) copies, from generally
different and unknown Gaussian noisy state preparation and without
any possibility to tailor the states going into the channel, our
task is to concentrate (deterministically or probabilistically)
information from the copies into a single copy with less noise.

A proposed basic two-copy purification scheme is depicted in Fig.~1.
The input two-mode density matrix is a tensor product of a pair of matrices (\ref{Gen01}).
Since this representation exploits a basis of two-mode coherent states,
we will study the evolution of the pure
state $|\alpha,\alpha'\rangle$ and use the results to obtain a final distilled state.
At the first balanced beam splitter the two input modes
interfere and produce a state $|\alpha,\alpha'\rangle \rightarrow
    |\frac{\alpha+\alpha'}{\sqrt{2}},\frac{\alpha-\alpha'}{\sqrt{2}}\rangle$.
The mode created by constructive interference is then passed
through an attenuator with a transmittance $T_0$
\begin{equation}\label{Sim03}
    |\frac{\alpha+\alpha'}{\sqrt{2}},\frac{\alpha-\alpha'}{\sqrt{2}},0\rangle \rightarrow
    |\frac{\sqrt{T_0}(\alpha+\alpha')}{\sqrt{2}},\frac{\alpha-\alpha'}{\sqrt{2}},\frac{\sqrt{1-T_0}(\alpha+\alpha')}{\sqrt{2}} \rangle,
\end{equation}
where the transmittance can, for any initial state (\ref{Gen01}),
be tuned to the value that leads to the most desirable outcome.We are interested in unity-gain purification to
preserve the signal encoded into the coherent states. Since the
noise model is considered to be additive, such unity-gain
purification is achieved for the value $T_0=0.5$ of the
attenuation parameter. The product of the distillation setup for
fixed states, the states $|\frac{\alpha+\alpha'}{2}\rangle_1$ and
$|\frac{\alpha+\alpha'}{2}\rangle_3$, is the same as in the
Gaussian procedure \cite{Ulrik}. That procedure, however, leaves
the remaining state $|\frac{\alpha-\alpha'}{\sqrt{2}}\rangle_2$
unmeasured.

An important step towards probabilistic purification for unknown
$N_1$ and $N_2$ is to integrate a binary type detector into the
setup; as is depicted in Fig.~1. Such a detector, that allows us to
discriminate between a signal and the vacuum, can be realized
either by an avalanche photo-diode (APD) or by a heterodyne
(homodyne) detection. In this probabilistic scheme, the result of
distillation is accepted if a detector, followed by appropriate
processing, affirms the vacuum state in port 2, the probability of
which shall be denoted as
$\mathcal{P}_.(\frac{\alpha-\alpha'}{\sqrt{2}})$, where the subscript $.$ stands
for different detectors. {A figure of merit of the quality of our
binary detector in discriminating the vacuum state from others is
the ratio
$\mathcal{R}_.(\gamma)=\mathcal{P}_.(\gamma)/\mathcal{P}_.(0)$. We
can say that the detector given by  $\mathcal{R}_1$ suits our task
better than the detector described by $\mathcal{R}_2$ if the
probability of post-selection of undesirable states is lower, that
is $\mathcal{R}_2(\alpha)>\mathcal{R}_1(\alpha)$ for all $\alpha
\neq 0$ } Similarly, mode 3 can be also measured, but in this case
both a measurement and the associated benefit strongly depend on a particular
form of the noise present in the system, and we will therefore
exclude mode 3 from our general analysis. Keep in mind, though,
that in certain scenarios the measurement on mode 3 can improve
our ability to control the purification. The complete evolution of
the considered two-mode coherent state can be then expressed as:
\begin{equation}\label{evolution}
|\alpha,\alpha'\rangle\langle \alpha,\alpha'|  \rightarrow
\mathcal{P}_.\left(\frac{\alpha-\alpha'}{\sqrt{2}}\right)|\frac{\alpha+\alpha'}{2}
\rangle\langle\frac{\alpha+\alpha'}{2}|\otimes
|\frac{\alpha+\alpha'}{2}\rangle\langle\frac{\alpha+\alpha'}{2}|.
\end{equation}

The initial state is expressed as a tensor product,
$\rho_{in}\otimes\rho_{in}$,  of a pair of the density matrices
(\ref{Gen01}) where the noise-introducing channels are generally
different. Now, by applying the relation  (\ref{evolution}) we
obtain the total state of the output modes
\begin{eqnarray}
\rho_{tot} &=& \frac{1}{S}\int\!\!\!\!\int
    \Phi_1(\beta_1)\Phi_2(\beta_2)
    \mathcal{P}_.(\frac{\beta_1-\beta_2}{\sqrt{2}})
    | \alpha_0 +\frac{\beta_1+\beta_2}{2}\rangle\langle \alpha_0
    +\frac{\beta_1+\beta_2}{2}|\otimes\nonumber\\
    & &| \alpha_0 +\frac{\beta_1+\beta_2}{2}\rangle\langle \alpha_0 +\frac{\beta_1+\beta_2}{2}|\dd^2\beta_1
    \dd^2\beta_2,
\end{eqnarray}
where $S$ is a normalization factor representing probability of
success. Thus, after the purification both previously uncorrelated copies become
classically correlated.
Due to the symmetry of the scheme both copies are identical after tracing over the
other one, and if they are treated independently it can lead to further noise reduction.
The particular state of the copy is
\begin{equation}\label{Gen02}
    \rho_{out} = \frac{1}{S}\int\!\!\!\!\int
    \Phi_1(\beta_1)\Phi_2(\beta_2)
    \mathcal{P}_.\left(\frac{\beta_1-\beta_2}{\sqrt{2}}\right)
    | \alpha_0 +\frac{\beta_1+\beta_2}{2}\rangle\langle \alpha_0 +\frac{\beta_1+\beta_2}{2}|\dd^2\beta_1
    \dd^2\beta_2,
\end{equation}
where $S$ denotes the probability of successful distillation and
can be found to be
\begin{equation}\label{Gen03}
    S = \int\!\!\!\!\int \Phi_1(\beta_1)\Phi_2(\beta_2)
    \mathcal{P}_.\left(\frac{\beta_1-\beta_2}{\sqrt{2}}\right) \dd^2\beta_1
    \dd^2\beta_2.
\end{equation}
From Eq.~(\ref{Gen02}), the symmetrizing property \cite{symm} in the
coherent-state purification is clear.  The noise amounts, arising in
particular channels, are averaged and symmetrized. If the
distributions $\Phi_1(\beta_1)$ and $\Phi_2(\beta_2)$ as well as the
filtration function
$\mathcal{P}_.(\frac{\beta_1-\beta_2}{\sqrt{2}})$ are Gaussian
functions of the arguments centered around the origin in phase space
then the output state is also Gaussian state.

To analyze the improvement we calculate the mean number of thermal
photons of the output state
\begin{equation}\label{Gen03a}
N'_.= \frac{\int\!\!\!\!\int
    \Phi_1(\beta_1)\Phi_2(\beta_2)
    \mathcal{R}(\frac{\beta_1-\beta_2}{\sqrt{2}})
    \left|\frac{\beta_1+\beta_2}{2}\right|^2
    \dd^2\beta_1
    \dd^2\beta_2}{\int\!\!\!\!\int \Phi_1(\beta_1)\Phi_2(\beta_2)
    \mathcal{R}(\frac{\beta_1-\beta_2}{\sqrt{2}}) \dd^2\beta_1
    \dd^2\beta_2.}
\end{equation}
and use it to compare obtained results. Other methods that we will
use in our comparison, are the deterministic purification (D)
\cite{Ulrik} and classical purification methods such as local
measurement of two copies with data averaging and subsequent
preparation (MP) and, eventually, with post selection according to
measured data (PMP). The deterministic Gaussian purification is
realized by letting two modes constructively interfere and then
attuning the final signal to achieve unity gain. The mean photon
number $N'_D$ can be straightforwardly obtained from (\ref{Gen03a})
by setting $\mathcal{P}_.\equiv 1$ and serves as an upper bound for
the probabilistic method with detector efficiency approaching zero.

\section{Detection and post-selection:}          

To calculate the $N'_.$ for a particular channel we will need a
proper expression for the post-selection probability
$\mathcal{P}_.(\alpha)$. One way to distinguish a vacuum state is
by an avalanche photo diode (APD). The action of a perfect APD can
be described by a pair of projection operators $\Pi_{\circ} =
|0\rangle\langle 0|$ and $\Pi_{\bullet} = 1-| 0\rangle\langle 0|$,
which correspond to measurement outcomes of 'no click' and
'click', respectively. The probability that an ideal APD will not
produce a click if the coherent state $|\alpha\rangle$ has
arrived, is $\mathcal{P}_{APD}(\alpha) = e^{-|\alpha|^2}$. An
imperfect detector can be modelled by a beam-splitter, with
transmissivity equal to the quantum efficiency $\eta_{APD}$, in
front of an ideal APD, and with a thermal state with mean number
of chaotic photons $n_{ch} = p_d/[(1-p_d)(1-\eta_{APD})]$ in a
free port of the beam-splitter, simulating dark counts with a rate
$p_{d}$ (probability of a dark count occurrence). The total
positive operator valued measure (POVM) for the imperfect
detector is
\begin{equation}
\Pi=1-(1-p_d)\sum_{n=0}^{\infty}(1-\eta_{APD}(1-p_d))^{n}|n\rangle\langle
n|.
\end{equation}
The no-click probability is then obtained as
\begin{equation}\label{Detectors}
    \mathcal{P}_{APD}'(\alpha) =(1-p_{d})\exp\left[-(1-p_{d})\eta_{APD}|\alpha|^2\right].
\end{equation}
Considering realistic common APDs having parameters
$\eta_{APD}\approx 0.4$ and $p_{d}\approx 10^{-4}$, the dark counts,
causing solely the reduction of the rate of the process, can be
omitted. However, the reduced efficiency $\eta_{APD}$ may, for the
weak coherent signals, lead to accepting wrong results and thus
increase the noise of the obtained state. For the usual low dark
count rate, $\mathcal{P}_{APD}(0)\approx 1$ and
$\mathcal{R}_{APD}\approx \exp(-\eta_{APD}|\alpha|^2)$, it is
evident that with increasing efficiency the ability of an APD, to
distinguish vacuum state from other coherent states, improves.

An alternative to a low efficiency avalanche photo-diode can be
found in the higher efficiency of heterodyne detection (eight-port
homodyne detection), which typically is $\eta_{HET}>0.95$. The
heterodyne detection measures simultaneously both the
complementary quadratures, $X$ and $P$ ($a = X+ \ii P$, $[X,P] = \ii/2$), of the
optical signal, after splitting the signal into two equally
intense parts. The high efficiency and low dark noise of homodyne
detectors arise from a sufficiently intense local oscillator.
Also, additive Gaussian noise in a local-oscillator channel is not a problem, in
balanced homodyne detection the fluctuations of the LO are completely suppressed
\cite{balanc}.
The need for a separate local oscillator can be
seen as a drawback, but, because the measured value $r$ is phase
insensitive, there is no need to maintain a phase-lock with the
signal (phase-randomized homodyne (heterodyne) detection
\cite{phasernd}), this moderately improves the situation.
It is
also noteworthy that the APD is a broadband detector, measuring an
entire spectrum of the signal, whereas the homodyne (heterodyne)
detection can be used to detect the signal encoded in a pair of
narrow-band frequency sidebands \cite{Ralph}. In frequency
multiplexed channels, many of these pairs of sidebands are
employed to carry the information. They are measured
simultaneously by a single homodyne (heterodyne) detection and
extracted by spectral analysis during the electronic processing.
Therefore, homodyne (heterodyne) detection can allow us to purify
the independent sideband channels simultaneously.

An ideal heterodyne measurement yielding complex result
$\gamma=\bar{x}+i\bar{p}$, where $\bar{x},\bar{p}$ are real measured
results of the quadratures $X,P$, is characterized by a POVM element
$|\gamma\rangle\langle \gamma|/\pi$, while imperfection is
represented by a virtual beam splitter, with transmissivity
$\eta_{HET}$, which is inserted into the signal's path. For a given
coherent state with amplitude $\alpha$, the imperfect heterodyne
measurement yields value $\gamma$
 with a probability given by a Q-function, $Q(\gamma) = |\langle \gamma|\eta_{HET}\alpha
\rangle|^2/\pi$. To approximate the work of an APD we use the
expansion $\gamma = r\exp(\ii \psi)$ and post-select the result if
the value of $r$ falls into the interval $\langle 0, R\rangle$.
The probability of post-selection is then obtained by integrating
the Q-function over this area
\begin{eqnarray}
\mathcal{P}_{HET}(\alpha) &=& \frac{2}{\pi}\int_{-\pi}^{\pi}\int_{0}^R r \exp\left( - r^2-\eta_{HET} |\alpha|^2 + 2r  \sqrt{\eta_{HET}}|\alpha|\cos\phi\right) \dd \phi \dd r
\end{eqnarray}
where $\phi$ is the relative phase difference between the phase of the
measured coherent state $|\alpha\rangle$ and a phase $\psi$.
By integrating over the variable $r$, one can directly get
\begin{eqnarray}
\mathcal{P}_{HET}(\alpha)&=&\frac{1}{2\pi}\exp\left( -\eta_{HET}|\alpha|^2\right)\times \nonumber\\
& &\int_{-\pi}^{\pi}\left\{
1-\exp\left[ (-R^2+2Ra )+\sqrt{\pi}a\ee^{a^2}\Bigl(\mbox{Erf}[a]- \mbox{Erf}[a-R]\Bigr)\right]\right\} \dd \phi,
\end{eqnarray}
where $a=|\alpha|\sqrt{\eta_{HET}}\cos\phi$. An interesting
property of heterodyne detection of coherent states becomes
apparent when $R$ tends to zero. Using
$\mathcal{R}_{HET}=\mathcal{P}_{HET}(\alpha)/\mathcal{P}_{HET}(0)$
to compare the binary detectors,  $\mathcal{R}_{HET}$ approaches
$\mathcal{R}_{HET}=\exp\left(-\eta_{HET}|\alpha|^2\right)$, at the
cost of decreasing success rate. This expression is identical to
that for an APD detection up to the detection efficiency. The
detector efficiency of heterodyne detection is usually
significantly greater than the efficiency of APDs. This can lead
to an improvement in fidelity if the rate of purification is not a
major issue.

Alternatively, a phase-randomized single quadrature homodyne detection could be considered as a possible detection method. In this situation, the signal is post-selected if the measured value falls within an interval $\langle -d,d\rangle$. The probability of post-selection is then
\begin{eqnarray}
\mathcal{P}_{HOM}(\alpha) &=&
\sqrt{\frac{2}{\pi}}\int_{-\pi}^{\pi}\int_{-d}^{d}
\exp\left(-2(x-|\alpha|\cos\phi)^2 \right) \dd \phi \dd x.
\end{eqnarray}
Calculating the ratio $\mathcal{R}_{HOM}$ in the limit $d\rightarrow 0$, we get
\begin{equation}
\mathcal{R}_{HOM}=\exp\left(-\eta_{HOM}|\alpha|^2\right)\mbox{J}_{0}\left[i\eta_{HOM}|\alpha|^2\right],
\end{equation}
where $J_{0}$ is the Bessel function of first order. Assuming that
heterodyne detection consists of two homodyne detectors with
the same efficiency, $\eta_{HOM}$, we have $\eta_{HET}=\eta_{HOM}$, and
we find that for any detectors having $\eta_{HOM}>0$,
$\mathcal{R}_{HOM}(\alpha)>\mathcal{R}_{HET}(\alpha)$ for
arbitrary $|\alpha|\not=0$. Thus, for the proposed purification, the
phase-randomized heterodyne detection is better than
phase-randomized homodyne detection.

However, if the noise is presented only in a single known quadrature
(for example, phase-quadrature $P$ ), one can, once
more, think about homodyne detection (not phase-randomized) as
an alternative for a distillation measurement. In this case,
the probability of the post-selection is
\begin{eqnarray}
\mathcal{P}_{HOM}(\alpha) &=& \sqrt{\frac{2}{\pi}}\int_{-d}^{d}
\exp\left(-2(p-|\alpha|\sin\theta)^2 \right) \dd p,
\end{eqnarray}
where $\alpha=|\alpha|\exp(i\theta)$ is a complex amplitude of the
coherent state, and the factor $\mathcal{R}_{HOM}(\alpha)$ is then
\begin{equation}
\mathcal{R}_{HOM}(\alpha)=\exp\left(-2\eta_{H0M}|\alpha|^2\sin^{2}\theta\right).
\end{equation}
If the homodyne detection can be locked to the quadrature suffering from noise then $\theta = \pi/2$ and the
homodyne detection with efficiency $\eta_{HOM}$ gives qualitatively the same results as heterodyne measurement
with $\eta_{HET} =\eta_{HOM}/2$. It is then obvious that, in case of noise presented in a single known quadrature,
the use of homodyne detection can be advantageous.

\section{Results:}            

Phase-insensitive excess noise is the most common additive noise
disturbing state preparation. The additive Gaussian excess noise in
the laser beam and in the modulators can be represented by as a
single source of noise in a classical-quantum channel described by
Eq.~(\ref{Gen00}) with a noise distribution
\begin{equation}\label{Gauss}
\Phi_{N}(\beta) = \frac{1}{\pi N}\exp(-\frac{|\beta|^2}{N}),
\end{equation}
where $N$ corresponds to the mean number of thermal photons
. The fidelity of coherent state preparation is then
given by $F'_.=1/(1+N'_.)$. Let us assume a simple
continuous-variable communication protocol with the coherent states
and heterodyne detection. The sender is preparing coherent states
from the prior Gaussian distribution having mean number $n$ of
signal photons and $N$ of thermal photons, and the receiver is using
the heterodyne detection to decode transmitted information. Optimal
detection in this case is heterodyne detection described by the POVM
$\Pi=\frac{1}{\pi}|\gamma\rangle\langle\gamma|$, where $\gamma$ is
the detected amplitude. The classical capacity of such communication
through narrow-band ideal channel was actually calculated in as
$C=\ln\left(1+n/(1+N)\right)$ \cite{Hall}. The capacity is a
monotonically decreasing function of the mean value of thermal
photons $N$.
A more deep impact has excess noise in the coherent state key
distribution protocol through lossy channel. An excess noise in the
trusted state preparation decreases secure key rate and can even
break security for a given attenuation of the channel
\cite{security1}.

There are two basic classical purification strategies based on
optimal measurement by heterodyne detection of every copy: data
processing and state re-preparation. If the actual $N_1$ and $N_2$
are not known, it is impossible to tailor the purification method
and it has to be symmetric with respect to the states. The first
method, already described in \cite{Ulrik}, reduces noise
deterministically by data averaging: measured results forming
complex numbers $\alpha_1$ and $\alpha_2$ are averaged,
$\alpha'=(\alpha_1+\alpha_2)/2$, and used to prepare a new coherent
state $|\alpha'\rangle$. The deterministic measurement-preparation
(MP) strategy results in a mean number of thermal photons
\begin{equation}\label{MP}
N'_{MP}=\frac{N_1+N_2}{4}+\frac{1}{2}
\end{equation}
in a single re-prepared copy.

To get some improvement, one could devise a classical filtering
scheme; re-preparing the signal only if the complex measured
values $\alpha_1$ and $\alpha_2$ satisfy
$|\alpha_1-\alpha_2|<\Delta$, where $\Delta$ is some small number
serving as a threshold. If we assume perfect heterodyne detection,
then as $\Delta$ tends to zero, the mean value of thermal photons can be found to satisfy
\begin{equation}\label{PMP}
\frac{1}{N'_{PMP}}=\frac{1}{1+N_1}+\frac{1}{1+N_2},
\end{equation}
at a cost ofically rapidly decreasing success rate. Since $N'_{PMP}\leq
N'_{MP}$ for all $N_1,N_2$, where equality occurs for $N_1=N_2$,
the probabilistic MP (PMP) method can improve the deterministic MP
method. For small mean photon numbers $N_1,N_2\ll 1$, the added
noise by this method approaches $N'_{PMP}\approx N_{MP}$. On the
other hand, if $N_1,N_2\gg 1$ then $N'_{PMP}\approx
N_1N_2/(N_1+N_2)$ and we can conditionally approach the noise
reduction corresponding to a case when both $N_1$ and $N_2$ are
precisely known, as can be seen below.

It is important to emphasized that any classical method (MP, PMP),
based on measurement and re-preparation, cannot be used in the state
preparation, because the new state would be again disturbed by the
same preparation noise. A method which can be applied for the state
preparation is the deterministic quantum purification protocol
\cite{Ulrik} based purely on the interference of the copies. Thus
the only important requirement for the application of the quantum
purification is the coherence between the copies leading to the
interference with high visibility, which can be achieved by using
standard quantum noise-locking techniques. A resulting number of
thermal photons can be obtained from (\ref{Gen03a}) by setting
$\mathcal{R}\equiv 1$ and it is
\begin{equation}\label{ch12G}
N'_{D}=\frac{N_1+N_2}{4}.
\end{equation}
Evidently this is better than the MP strategy, but comparing
(\ref{ch12G}) and (\ref{PMP}), $N_{PMP}>N_{D}$ only if
\begin{equation}
\frac{(N_1-N_2)^2}{2(N_1+N_2)+4}<1.
\end{equation}
Thus especially for highly asymmetrical channels with large total
mean photon number $N_1+N_2$, $N'_{PMP}$ can be substantially lower
than $N'_{D}$.

Note, if there is a possibility of estimating $N_1$ and $N_2$, the
symmetrical deterministic protocol can be tailored to achieve the best
performance. If both the beam splitter $BS$ and attenuator with
the transmissivity $T$ and $T_0$, respectively, are properly adjusted as
\begin{equation}
T=\frac{N_{2}^{2}}{N_{1}^{2}+N_2^2},\,\,\,T_0=\frac{N_{1}^{2}+N_2^2}{(N_1+N_2)^2},
\end{equation}
then one can approach the following reduction of noise excess:
\begin{equation}
\frac{1}{N'_{T}}=\frac{1}{N_1}+\frac{1}{N_2}.
\end{equation}
As such the protocol is still Gaussian and completely
deterministic.

In the case where $N_1$ and $N_2$ are unknown, the deterministic
method can be overcome by a probabilistic strategy, at a cost of the
preparation rate. If we consider the purification scheme as on
Fig.~1, with the APD having detection efficiency $\eta_{APD}$ in
mode $2$, the resulting mean photon number is given by
(\ref{Gen03a}) with (\ref{Detectors}) and satisfies
\begin{equation}\label{ch12NG}
\frac{1}{N'_{APD}+\frac{1}{2\eta_{APD}}}=\frac{1}{\frac{1}{\eta_{APD}}+N_1}+\frac{1}{\frac{1}{\eta_{APD}}+N_2}.
\end{equation}
The distillation will succeed with a probability
\begin{equation}
S=\frac{2}{2+\eta_{APD}(N_1+N_2)}.
\end{equation}
Comparison of (\ref{ch12NG}) with (\ref{ch12G}) yields that, for
arbitrary $\eta_{APD}>0$, the probabilistic purification always
beats the deterministic purification as long as $N_1 \neq N_2$. For
$N_1=N_2$ both the methods give the same result as the deterministic
method $N'_{APD}=N'_{D}=N/2$, independently on $\eta_{APD}$. Note,
to overcome $PMP$ method for any $N_1$ and $N_2$
($N'_{PMP}>N'_{APD}$), it is necessary to use an APD with
$\eta_{APD}>1/2$. But then we get a better noise reduction with
finite probability of success, not only asymptotically as for the
$PMP$ method. Thus, for a pair of Gaussian channels characterized by
unknown, mean chaotic photon numbers $N_1$ and $N_2$, of all the
methods we considered, the probabilistic purification with an APD
leads to the best result. However, it is also possible to implement
such measurements using heterodyne detection and benefit from its
higher efficiency, as has been discussed in the previous section. It
is an interesting result since for our task we can substitute
detection of vacuum state using the APD by the heterodyne detection,
if a lower success rate is accepted. If unit efficiency
$\eta_{APD,HET}=1$ is approached, the mean photon number from this
method is simply $N'_{APD}=N'_{PMP}-1/2$.

In Fig.~2, the mean number $N'_{APD}$ of thermal photons and success
rate $S$ are plotted against $N_1$ and $N_2$ for $\eta_{APD}=1$. For
comparison, Fig.~3 shows behavior of $N'_D$ and $N'_T$. We can
observe that, for a weak noise $N_1,N_2\ll 1$, an improvement of the
probabilistic method over the deterministic method is only moderate,
because $N'_{APD}$ approaches $N'_{D}$. On the other hand, for
$N_1,N_2\gg 1$, the mean photon number $N'_{APD}$ approaches
$N'_{T}$, that is, the noise reduction is almost as good as in the
case when Clare precisely knows $N_1$ and $N_2$. The tailored
deterministic method, based on precise knowledge of $N_1$ and $N_2$,
will always surpass the probabilistic method, but in the limit of
large $N_1+N_2$ and strongly asymmetric channels, the difference in
mean numbers of thermal photons
\begin{equation}
N_{APD}-N_{T}=\frac{(N_1-N_2)^2}{2(N_1+N_2)(N_1+N_2+2)},
\end{equation}
approaches a constant value of $1/2$.

\begin{figure} [htbp]
\centerline{\psfig{file=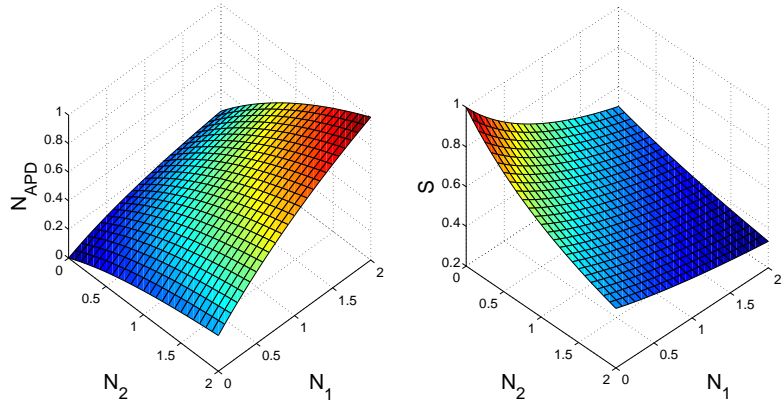,width=12cm}}
\caption{\label{result} The mean number $N'_{APD}$
of thermal photons and success rate $S$ of the purification after
the non-Gaussian purification with ideal APD detector
($\eta_{APD}=1$) as a function of mean numbers $N_1$ and $N_2$ of
thermal photons in the channels.}
\end{figure}
\begin{figure} [htbp]
\centerline{\psfig{file=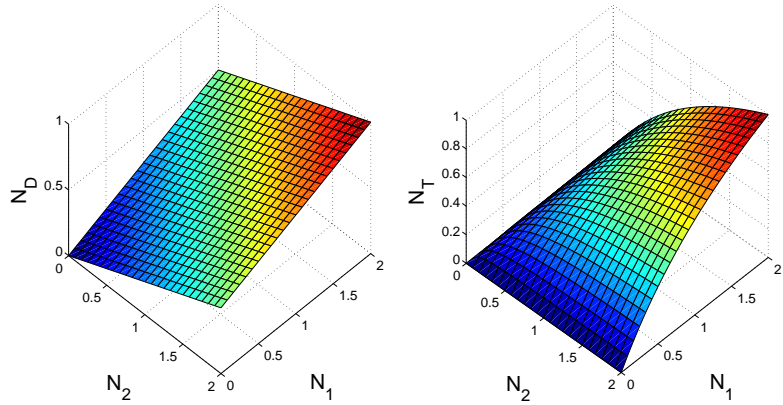,width=12cm}}
\caption{\label{result2} The mean numbers $N'_{D}$ and $N'_{T}$
of thermal photons as a function of mean numbers $N_1$ and $N_2$ of
thermal photons in the channels.}
\end{figure}

Let us now consider another example. If the coherent state is
encoded by a relatively weak modulation of a bright carrier beam
with fixed amplitude but exhibiting phase fluctuation, a different
type of noise can occur. The noise distribution is
\begin{equation}\label{GaussP}
\Phi_{N}(\beta) = \frac{1}{\sqrt{2\pi N}}\exp(-\frac{\beta_{I}^2}{2N})\delta(\beta_{R}),
\end{equation}
with $\beta=\beta_R+\ii\beta_I$, and corresponds to perfect
transmission in one quadrature (amplitude) and Gaussian fluctuation
in the other (phase). The $N$ stands again for the mean number of
thermal photons added in the state preparation. For phase
fluctuations which are weak relative to the amplitude of the
carrier, the amplitude quadrature $X$ will remain uninfluenced, and
only additive noise will be introduced in the phase quadrature $P$.
Therefore, we are going to be interested only in the improvement of
the $P$ quadrature and so make the reasonable request that
distillation should not add any noise into the quadrature $X$. Note
that this demand cannot be satisfied by any method utilizing
measurement and re-preparation.

As well as for the phase-sensitive noise, the deterministic Gaussian
method gives $N'_D = (N_1+N_2)/4$. As in the previous case, if
$N_1=N_2$, then nothing can be gained by the probabilistic
protocol.
However, if, in general, $N_1 \neq N_2$,the use of an APD (or
heterodyne detection) can reduce the mean number of chaotic
photons to
\begin{equation}\label{phase1}
\frac{1}{N'_{APD}+\frac{1}{4\eta_{APD}}}=
\frac{1}{\frac{1}{2\eta_{APD}}+N_1}+\frac{1}{\frac{1}{2\eta_{APD}}+N_2}.
\end{equation}
Furthermore, since we have assumed the noise to occur only in a single known quadrature, we
can implement another type of measurement. If we decide to measure the $P$ quadrature by
homodyne detection and post-select the signal only if the detected value falls into an
interval $\langle -d,d\rangle$, then as $d$ tends to $0$, the mean number of chaotic photons in the
purified state approaches
\begin{equation}\label{phase2}
\frac{1}{N'_{HOM}+\frac{1}{8\eta_{HOM}}}=\frac{1}{\frac{1}{4\eta_{HOM}}+N_1}+
\frac{1}{\frac{1}{4\eta_{HOM}}+N_2}.
\end{equation}
By comparing (\ref{phase1}) and (\ref{phase2}) we can see that, if
we are able to provide a phase-locked local oscillator to perform
proper homodyne detection, we may benefit from double the
efficiency of phase-randomized heterodyne measurement, implemented
by the same detectors. Therefore, for any $\eta>0$, we have
$N'_{HOM} \leq N'_{APD} \leq N'_{D}$, where the $\eta_{APD} =
\eta_{HOM} = \eta$. The equality holds only for $N_1 = N_2$.

Comparing this method (with ideal detector) with the tailored
deterministic method, resulting in $N'_T=N_1N_2/(N_1+N_2)$, the difference of the photon numbers
\begin{equation}
N_{HOM}-N_{T}=\frac{(N_1-N_2)^2}{8(N_1+N_2)\left(\frac{1}{2}+N_1+N_2\right)}
\end{equation}
behaves similarly as for the phase insensitive noise discussed
above. That is, in limiting cases $N_1,N_2\ll 1$ ($N_1,N_2\gg 1$),
the mean photon number  approaches $N_{HOM}\approx N_{D}$
($N_{HOM}\approx N_{T}$).

\begin{figure} [htbp]
\centerline{\psfig{file=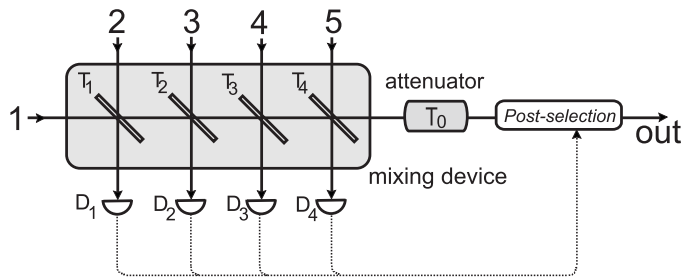, width=9cm}} 
\caption{\label{setup2} A scheme for probabilistic
purification of coherent states (multi-copy purification, $M=5$):
D$_1$-D$_4$ -- avalanche photodiodes or heterodyne (homodyne)
detection.}
\end{figure}

The probabilistic purification scheme can be straightforwardly
extended for setups involving a greater number of copies, as is
schematically depicted at Fig.~4. The input modes can be combined at
an array of the beam splitters with $T_j=(j-1)/j$, $j=2,\ldots,M$,
and all outputs, except the one where the constructive interference
occurs, are detected by the APDs (or heterodyne detections). The
final state is only accepted if all the detectors confirm zero
signal. Then the output with constructive interference is properly
attenuated ($T_0=1/M$) to achieve the unity gain regime. Similarly
as in the two-copy case, $M$ classically correlated copies with
reduced noise are actually produced.

Using $M$ copies of noisy state with mean numbers of thermal photons
$N_1,\ldots,N_M$, the deterministic purification leads to an output
state with mean value of thermal photons
\begin{equation}
N'_{D}=\frac{1}{M^2}\sum_{i=1}^{M}N_i.
\end{equation}
The result of the probabilistic method ($\eta_{APD}=1$) can be
expressed as
\begin{equation}
\frac{1}{N'_{APD}+\frac{1}{M}}=\sum_{i=1}^{M}\frac{1}{1+N_i}
\end{equation}
with the probability of success
\begin{equation}
S=\frac{M}{M+(M-1)\sum_{i=1}^{M}N_i+(M-2)\sum_{i=1}^{M}N_{i}N_{j\not=i}+(M-3)
\sum_{i=1}^{M}N_{i}N_{j\not=i}N_{k\not=i,j}+\ldots}.
\end{equation}
If $N_1,\ldots,N_M$ are known, it is again possible to tailor the
transmissivities $T_i$ in the deterministic purification and achieve
$N'_{T}$ given by:
\begin{equation}
\frac{1}{N'_{T}}=\sum_{i=1}^{M}\frac{1}{N_i}.
\end{equation}
As in the two-mode case, if $N_1,\ldots,N_M\ll 1$, $N'_{APD}$
approaches $N'_{D}$ and it is sufficient to use the deterministic
method. On the other hand, for $N_1,\ldots,N_M\gg 1$, the $N'_{APD}$
approaches $N'_{T}$ and the probabilistic method can lead to noise
reduction almost at the level of perfect knowledge, if preparation
rate is sacrificed. In the limit of large $M$, the probabilistic
quantum method approaches
\begin{equation}
\frac{1}{N'_{APD}}=\sum_{i=1}^{M}\frac{1}{1+N_i}
\end{equation}
independently of values of $N_i$. Similar results and discussion can
be analogously performed for phase-sensitive noise.

\section{Conclusion:}            

In summary, we have demonstrated a feasible probabilistic
purification method can reduce Gaussian additive excess noise noise
in the coherent-state preparation and overcome previous
deterministic method \cite{Ulrik}. Since the excess noise can be
unstable and its actual level can be unknown we extended original
idea to such the realistic case. Based on previously experimentally
tested deterministic purification of coherent states \cite{Ulrik},
the method relies on using interference of two noisy modes on a
balanced beam splitter and post selecting one of the modes (BS
output with constructive interference) if there is no signal from
the avalanche photo-diode (APD) measuring the other mode (BS output
with destructive interference). It was also shown that heterodyne
detection (approaching unit efficiency) can be used instead of the
APD, if reduction in transmission rate can be accepted. Also, for
the phase sensitive noise, post-selection according to the homodyne
detection can reduce the noise even further. An extension of the
scheme arbitrary number of noisy copies is presented. It has a
direct application in an improving classical capacity of the
coherent-state communication. Since the trusted state preparation is
assumed to be under full control of the sender, the proposed quantum
purification can be used to reduce excess noise in the CV secure key
distribution \cite{CVcrypt}.

\medskip
\noindent {\bf Acknowledgments} The work was supported by the
project 202/03/D239 of Grant Agency of Czech Republic, project
MSM6198959213 of Czech Ministry of Education and EU project
FP6-511004 COVAQIAL. R.F. thanks the Alexander von Humboldt
fellowship and P.M. acknowledges support from the European Social
Fund.

\end{document}